\documentclass[twocolumn,prb,superscriptaddress]{revtex4}

\usepackage{amssymb}
\usepackage{amsmath}
\usepackage{graphicx}

\begin{document}

\title{The annealing mechanism of AuGe/Ni/Au ohmic contacts to a two-dimensional electron gas in GaAs/$\boldsymbol{{\rm Al}_{x}{\rm Ga}_{1-x}{\rm As}}$ heterostructures}

\author{E.~J.~Koop}
\author{M.~J.~Iqbal}
\author{F.~Limbach}
\author{M.~Boute}
\author{B.~J.~van~Wees}
\affiliation{Zernike Institute for Advanced Materials, Nijenborgh 4, University of
Groningen, NL-9747AG Groningen, The Netherlands}
\author{D.~Reuter}
\author{A.~D.~Wieck}
\affiliation{Angewandte Festk\"{o}rperphysik, Ruhr-Universit\"{a}t
Bochum, D-44780 Bochum, Germany}
\author{B.~J.~Kooi}
\author{C.~H.~van~der~Wal}
\affiliation{Zernike Institute for Advanced Materials, Nijenborgh 4, University of
Groningen, NL-9747AG Groningen, The Netherlands}

\date{\today}

\begin{abstract}
Ohmic contacts to a two-dimensional electron gas (2DEG) in
GaAs/${\rm Al}_{x}{\rm Ga}_{1-x}{\rm As}$ heterostructures
are often realized by annealing of AuGe/Ni/Au that is
deposited on its surface. We studied how the quality of
this type of ohmic contact depends on the annealing time
and temperature, and how optimal parameters depend on the
depth of the 2DEG below the surface. Combined with
transmission electron microscopy and energy-dispersive
\mbox{X-ray} spectrometry studies of the annealed contacts,
our results allow for identifying the annealing mechanism
and proposing a model that can predict optimal annealing
parameters for a certain heterostructure.
\end{abstract}


\maketitle


\section{\label{sec:intro}Introduction}

Epitaxially grown GaAs/${\rm Al}_{x}{\rm Ga}_{1-x}{\rm As}$
heterostructures that contain a two-dimensional electron
gas (2DEG) are widely used for electron transport studies
in low-dimensional systems
\cite{Beenakker1991,Kouwenhoven1998}. Establishing
electrical contacts to the 2DEG is a crucial step in device
fabrication with these heterostructures. A commonly used
recipe for making ohmic contacts is annealing of a
AuGe/Ni/Au alloy that has been deposited on the
heterostructure surface \cite{Baca1997}. High-quality
heterostructures are often only available in a limited
quantity, and it is desirable to minimize the heating that
is needed for annealing the contacts to avoid damaging the
heterostructure. A model that predicts optimal annealing
times and temperatures for a heterostructure with the 2DEG
at a certain depth is therefore very valuable.

We present here a study of the annealing mechanism for this
type of ohmic contact, and a model that can predict optimal
annealing parameters for a certain heterostructure. We used
electron transport experiments to study how the quality of
AuGe/Ni/Au based ohmic contacts depends on annealing time
and temperature, and how the optimal parameters change with
the depth of the 2DEG below the surface. These results
confirm that the annealing mechanism cannot be described by
a single simple diffusion process. Cross-sectional studies
of annealed contacts with Transmission Electron Microscope
(TEM) and Energy Dispersive \mbox{X-ray} (EDX) techniques
were used for identifying a more complex annealing
mechanism, that is in agreement with the results from our
electron transport studies.

The AuGe/Ni/Au contact was first introduced by Braslau
\textit{et al.} \cite{Braslau1967} to contact
\mbox{n-GaAs}, and several studies aimed at understanding
the contact mechanism for this type of contact
\cite{Ogawa1980,Braslau1981,Lee1981,Heiblum1982,Kuan1983,Braslau1983,Braslau1986,Procop1987,Waldrop1987,Bruce1987,Shappirio1987,Relling1988,Weizer1988,Lumpkin1996}.
Later studies focussed on the formation of an ohmic contact
to a 2DEG in a GaAs/${\rm Al}_{x}{\rm Ga}_{1-x}{\rm As}$
heterostructure
\cite{Zwicknagl1986,Higman1986,Rai1988,Jin1991,Taylor1994,Messica1995,Taylor1998,Raiser2005,Saravanan2008},
but do not report how the optimal annealing parameters
depend on the depth of the 2DEG below the surface. A number
of these studies suggest that a contact is formed because a
pattern of Au/Ni/Ge spikes that originate from the
metallization penetrate the heterostructure, just beyond
the depth of the 2DEG \cite{Taylor1994,Taylor1998}. Earlier
work had already suggested that in good contacts Ni-rich
phases may form at the depth of the 2DEG, in contact with
the GaAs below the ${\rm Al}_{x}{\rm Ga}_{1-x}{\rm As}$
\cite{Zwicknagl1986}. We observe, instead, a mechanism
where metal-rich phases only penetrate the heterostructure
over a distance that is shorter than the depth of the 2DEG.
The mechanism that results in a good contact is then
similar to a process that has been described
\cite{Kuan1983} for contacts to \mbox{n-GaAs}: during
annealing, the AuGe/Ni/Au on the surface segregates in
Ni-rich and Au-rich domains, where the Ni domains contain
most of the Ge. These domains penetrate the heterostructure
and grow towards the 2DEG in large grains rather than
narrow spikes. For optimal electrical contact conditions
the Au and Ni-rich grains do not reach the 2DEG. The
contact resistance decreases and the contact becomes ohmic
because Ge diffuses deeper, forming a highly doped ${\rm
Al}_{x}{\rm Ga}_{1-x}{\rm As}$ region between the 2DEG
layer and metal-rich phases at the surface. We find that
even for very long annealing times, when the contact
resistance has significantly increased compared to the
optimal contact, the Au and Ni-rich phases still do not
penetrate the 2DEG.

The outline of this article is as follows: we first
describe our wafer materials and device fabrication. Next,
we present electrical measurements and use these to
identify annealed contacts with optimal ohmic properties.
In Section~\ref{sec:temedx} we present the results of our
TEM and EDX studies of annealed contacts.
Section~\ref{sec:annmech} then summarizes the annealing
mechanism that we identified, and this is used in
Section~\ref{sec:model} to propose a model that can predict
optimal annealing parameters. Finally, in
Section~\ref{sec:shape}, we present a study of how the
contact resistance depends on the shape of the contact
(varying area or circumference), which gives further
insight in the annealing mechanism and the electrical
contact properties.


\section{\label{sec:fab}Device fabrication}

We studied annealed AuGe/Ni/Au contacts to three GaAs/${\rm
Al}_{x}{\rm Ga}_{1-x}{\rm As}$ heterostructures, grown on
(001)-oriented i-GaAs substrates, with the 2DEG at a
heterojunction at 70 nm (wafer~A), 114 nm (wafer~B), and
180 nm (wafer~C) below the surface of the wafer. These
wafers have similar values for the 2DEG electron density
$n_{s}$ and mobility $\mu$ (around $ 2 \cdot 10^{15} \:
{\rm m^{-2}} $ and $ 100 \: {\rm m^{2}/Vs}$, respectively,
results for 4.2~K and samples kept in the dark during cool
down). For all three wafers the layer sequence (from the
surface down) is very similar besides the depth of the
2DEG. The top layer is a $\sim5$~nm \mbox{n-GaAs} capping
layer, then an ${\rm Al}_{x}{\rm Ga}_{1-x}{\rm As}$ doping
layer (Si at $\sim1 \cdot 10^{18} \: {\rm cm^{-3}} $) with
$x \approx 0.32$, of thickness 30~nm (A), 72~nm (B) or
140~nm (C). After this follows an undoped ${\rm Al}_{x}{\rm
Ga}_{1-x}{\rm As}$ buffer layer ($\sim$35~nm thick). The
2DEG is located at the interface with the next layer, which
is a several $\mu$m thick undoped GaAs layer.

We studied 200 $\times$ 200 ${\rm \mu m^2}$ contacts that
were defined by optical lithography on a 1~mm wide and 2~mm
long etched mesa with a typical Hall-bar geometry. An
electron-beam evaporator was used for deposition of
subsequently 150 nm AuGe of eutectic composition (12 wt$\%$
Ge), 30 nm of Ni and 20 nm of Au. Subsequent annealing took
place in a pre-heated quartz furnace tube with a clean
$N_2$ flow of about 1~cm/s over the sample to prevent
oxidation. We also found that using a much weaker $N_2$
flow could result in surface contamination that was
electrically conducting. We have used three different
annealing temperatures, 400~$^{\circ}$C, 450~$^{\circ}$C
and 500~$^{\circ}$C. The samples were placed on a quartz
boat and then moved into the center of the oven for various
annealing times. Figure \ref{Fig:TimeTempDepth}a shows the
temperature of the surface of the quartz boat as a function
of time for an oven temperature of 450~$^{\circ}$C. We
assume that the sample temperature closely follows the
temperature of the quartz boat, since we assured a good
thermal contact over the full surface of the sample.


\section{\label{sec:elecmeas}Electrical measurements}

We measured the current-voltage (IV) characteristics of all
contacts to determine optimal annealing parameters. We
found that a suitable and sufficient definition for an
optimal ohmic contact is a contact with the lowest
zero-bias resistance at 4.2~K. The typical resistance for
such a contact is $\sim$20 $\Omega$, but we have observed
resistances as low as 5 $\Omega$. These values for contact
resistance are close to the lowest values that have been
reported \cite{Rvalues}. All contacts defined as optimal in
this manner showed highly linear IVs up to at least 1 mV
(over and under annealed contacts did show non-linear IVs
due to effects like Schottky or tunnel barriers in the
contacts). Further, all these optimal contacts showed a
strong monotonous reduction of the contact resistance upon
lowering the sample temperature from 300~K to 4.2~K. Highly
over and under annealed contacts showed an increase of the
contact resistance upon cooling to 4.2~K.

\begin{figure}[h!]
\includegraphics[width=1.0\columnwidth]{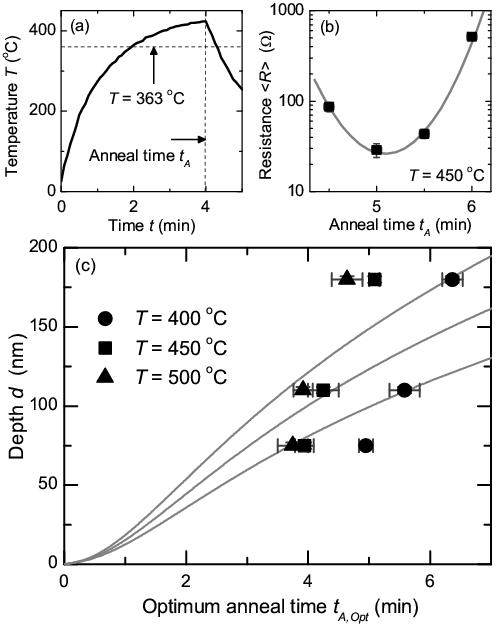}

\caption{(a) Temperature of the quartz boat as a function
of time for an oven temperature of 450~$^{\circ}$C.
Horizontal dashed line indicates the AuGe melting
temperature $T$ = 363~$^{\circ}$C. The vertical dashed line
indicates our definition of the annealing time $t_A$, the
time at which the boat is taken out of the oven. (b)
Average contact resistance $\langle R \rangle$ as a
function of annealing time $t_A$ for contacts on wafer~C,
annealed at 450~$^{\circ}$C. A parabolic fit is made to
estimate the annealing time where the resistance has a
minimum. (c) Overview of optimal annealing times
$t_{A,Opt}$ as a function of depth $d$ of the 2DEG beneath
the wafer surface for $T$ = 400, 450 and 500~$^{\circ}$C.
The three gray lines (bottom to top for 400, 450 and
500~$^{\circ}$C) are results of fitting a simple diffusion
model to the experimental data (see text), which does not
yield a good fit.} \label{Fig:TimeTempDepth}
\end{figure}

We used a current-biased 4-terminal configuration to
measure the voltage drop across a single contact, with two
terminals attached to the metal bond wire that is pressed
onto the ohmic contact, and two terminals attached to the
2DEG via other ohmic contacts (this allowed us to use a
standard sample design in our fabrication facility). We
accounted for a small voltage drop in the 2DEG area between
the contact and the voltage probe via the 2DEG. We are
aware that the Transmission Line Method (TLM)
\cite{Berger1972,Reeves1982} is a better method for
determining the exact value of a contact resistance, but
this is not needed for our approach. We compare resistances
of various annealed contacts that were fabricated under
identical conditions besides the variation in annealing
time and temperature. Within such a set, we determine which
contacts have the lowest contact resistance. When
reproducing our results with contacts that were fabricated
in a different batch (using the same electron-beam
evaporator, but after replenishing the AuGe target), we
find that the values of the lowest contact resistance can
be different up to a factor 2 around the typical result. We
attribute these batch-to-batch fluctuations to variations
in the exact composition of the AuGe/Ni/Au layer that we
deposit. The optimal annealing times, however, show
batch-to-batch fluctuations of only 10\%. Thus, our
approach for determining optimal annealing conditions does
not depend on the exact value of the measured contact
resistance.

Figure \ref{Fig:TimeTempDepth}b shows a typical result,
from which we determine the optimal annealing time for
contacts to wafer~C for the case of annealing with the oven
at 450~$^{\circ}$C. Contact resistance data that is denoted
as $\langle R \rangle$ is the average resistance measured
on a set of 8 identical contacts, and the error bar
represents the standard deviation. The results in
Fig.~\ref{Fig:TimeTempDepth}b show a clear minimum in
contact resistance for annealing times near 5 minutes. We
fit a parabola (phenomenological ansatz) to the $\log
\langle R \rangle$ values of these data points, and define
the optimal annealing time as the time coordinate of the
minimum of the parabola. In this manner, the optimal
annealing times $t_{A,Opt}$ are obtained for contacts on
wafers~A, B and C, annealed at each of the temperatures.


Figure \ref{Fig:TimeTempDepth}c presents these optimal annealing
times. As expected, the optimal annealing time increases as the
temperature is decreased, and increases as the depth $d$ of the 2DEG
increases. While it is known that several simultaneous diffusion
processes play a role in contact formation \cite{Kuan1983}, we
will, for the sake of argument, show that a simple diffusion model
has little value for predicting how optimal annealing times depend
on the depth $d$ and the annealing temperature. For this simple
diffusion model, we assume that a certain dopant (with fixed
concentration $C_0$ at the surface) diffuses into the
heterostructure. The relevant solution to Fick's second law is then
      \begin{equation} \label{Eq:FickDiffusion}
           C = C_0~{\rm erfc} \frac{z}{\sqrt{4Dt}} .
      \end{equation}
Here $C$ is the doping concentration at time $t$ and depth
$z$ into the heterostructure, and $D$ is the diffusion
constant (${\rm erfc}$ is complementary error function).
Since the temperature of our sample is not constant (see
Fig. \ref{Fig:TimeTempDepth}a) we will use the measured
temperature profile $T(t)$ to integrate the diffusion
constant over time, and use in Eq.~\ref{Eq:FickDiffusion}
$\int D(t)dt$ instead of $Dt$, where
      \begin{equation} \label{Eq:DiffusionConst}
           D(t) = D_0~{\rm exp}(-\frac{E_a}{k_B T(t)}),
      \end{equation}
and where $E_a$ is an activation energy. We assume that an
optimal contact then always occurs for a certain optimal
value for $C/C_0$ at the depth of the 2DEG ($z=d$). We
define the annealing time as the time from start to the
moment when the boat is taken out of the oven, but
integrate over the entire time span that the sample is at
elevated temperatures, (as shown in
Fig.~\ref{Fig:TimeTempDepth}a, fully including the cooling
down). This gives a model with the activation energy $E_a$,
diffusion constant $D_0$ and concentration $C/C_0$ as
fitting parameters.

The gray lines in Fig.~\ref{Fig:TimeTempDepth}c show the
best fitting result that reasonably covers all 9 data
points in a single fit. Besides the fact that the shape of
the traces only poorly matches the trend in the data, the
parameter values give unreasonable results. The temperature
dependence alone governs the fitting result for $E_a$,
giving here 0.15~eV. This is on the low side for typical
values for diffusion in GaAs materials ($\sim$1~eV)
\cite{Sarma1984,Kulkarni1988a,Kulkarni1988b}. For fixed
$E_a$, various combinations of $C/C_0$ and $D_0$ give
identical results. When assuming a typical value
$D_0\sim3\cdot 10^{-7} \; {\rm m^2/s}$ (for diffusion of
Ge, Ni or Au in GaAs
\cite{Sarma1984,Kulkarni1988a,Kulkarni1988b}), this fit
yields $C/C_0$ very close to 1, i.e. completely saturated
diffusion. This is in contradiction with the clear
dependence on depth that we observe (and this remains the
case when allowing for $E_a$ up to $\sim1$~eV, but then the
fit does not cover all 9 data points at all). Thus, we find
that predicting optimal annealing times with simple
diffusion (according to $t_{A,Opt} \propto d^2$ at constant
temperature) does not work and that a more complex model
needs to be considered.


\section{\label{sec:temedx}TEM and EDX results}

We have studied the contact formation using cross-sectional
TEM imaging of contacts at several stages during the
annealing process. The samples were prepared for TEM
imaging by using a Focussed Ion Beam (FIB) to slice out a
micrometer thin piece of the measured contact. By further
thinning using the FIB the thickness was reduced to 100~nm.


\begin{figure}[h!]
\includegraphics[width=1.0\columnwidth]{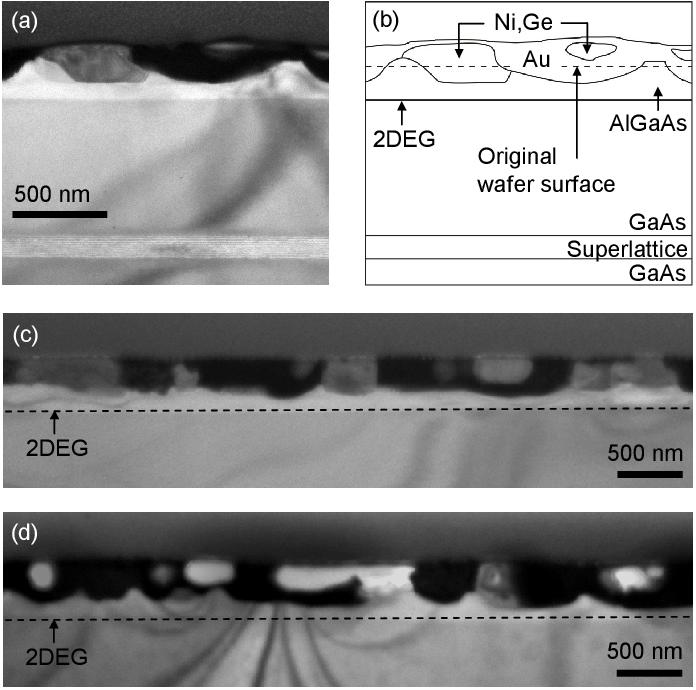}
\caption{(a) Cross-section TEM image of a contact on
wafer~C, annealed for the optimal annealing time at
450~$^{\circ}$C. (b) A sketch of the TEM image in (a) to
specify the various layers and phases. (c) Larger area TEM
image of the same contact as in (a) showing large Au-rich
(black) and Ni-rich grains (dark gray) contacting the ${\rm
Al}_{x}{\rm Ga}_{1-x}{\rm As}$. (d) Similar image for a
highly over annealed contact. The Au and Ni grains still do
not penetrate the 2DEG, but Au has diffused underneath the
Ni grains, which results in an increased contact
resistance.} \label{Fig:TEMimaging}
\end{figure}

Figure~\ref{Fig:TEMimaging}a shows an overview of an
optimally annealed contact on wafer~C which was annealed
for 5 minutes at 450~$^{\circ}$C. The composition of the
various phases has been determined by Energy Dispersive
\mbox{X-ray} (EDX) analysis and is illustrated in
Fig.~\ref{Fig:TEMimaging}b. From bottom to top we recognize
the GaAs substrate, and an AlAs/GaAs superlattice to
smoothen the surface of the substrate. On top of that we
find another layer of epitaxially grown GaAs and a layer of
${\rm Al}_{x}{\rm Ga}_{1-x}{\rm As}$. The 2DEG is at the
interface of these two layers. The GaAs capping layer that
was originally on top of the ${\rm Al}_{x}{\rm
Ga}_{1-x}{\rm As}$ layer is no longer visible. Instead, we
see large Au-rich and Ni-rich grains that have penetrated
below the original wafer surface. Both of these phases
contain out-diffused Ga and As, with Ga mainly in the
Au-rich grains and As mainly in the Ni-rich grains.
Further, the Ni-rich phase absorbed most of the Ge. We find
that the Au grains do not contain any Ge, consistent with
the findings of Kuan \textit{et al.} \cite{Kuan1983} in
work with \mbox{n-GaAs}.

The wide and curved dark lines going over all the
heterostructure layers (most clearly visible in the GaAs
layers) are due to strain induced by the FIB sample
preparation and are not related to the diffusion process.

We find that the Au-rich and Ni-rich grains do not have to
penetrate the 2DEG in order to establish a good electrical
contact. We can rule out that we do not see grains reaching
the 2DEG due to the small thickness of the sample slice,
since we observed no substantial variation in the
penetration depth of a large number of Au and Ni grains
going along the sample slice. We examined two slices from
two different samples, both with a length of 100~${\rm \mu
m}$, after electrical measurements confirmed that these
contacts were indeed optimally annealed.

The TEM image in Fig.~\ref{Fig:TEMimaging}c shows a larger
region of an optimally annealed contact. Large Au and Ni
grains that have penetrated the ${\rm Al}_{x}{\rm
Ga}_{1-x}{\rm As}$ layer can be identified.
Figure~\ref{Fig:TEMimaging}d shows an over annealed contact
on wafer~C, that was annealed for 7 minutes at
450~$^{\circ}$C. Remarkably, the Au and Ni grains did not
penetrate much further into the ${\rm Al}_{x}{\rm
Ga}_{1-x}{\rm As}$, and do still not reach the 2DEG. The
most significant change with respect to
Fig.~\ref{Fig:TEMimaging}c is that the Au-rich phase is
diffusing underneath the Ni-rich grains, reducing the total
Ni-grain--${\rm Al}_{x}{\rm Ga}_{1-x}{\rm As}$ interface
area. This was also observed by Kuan \textit{et al.}
\cite{Kuan1983} (and confirmed in detailed studies by Lumpkin
\textit{et al.} \cite{Lumpkin1996}) in work on
\mbox{n-GaAs}, and the results of these authors indicate
that this process is mainly responsible for the increase in
contact resistance when a sample is being over annealed.

Kuan \textit{et al.} \cite{Kuan1983} report that the
contact resistance is sensitive to the ratio of the total
contact area between Au-rich regions and ${\rm Al}_{x}{\rm
Ga}_{1-x}{\rm As}$, and that of Ni-rich regions. The
Au--${\rm Al}_{x}{\rm Ga}_{1-x}{\rm As}$ interface is
considered a region of poor conductance because the Au-rich
grains (in contrast to Ni-rich grains) do not contain any
Ge, such that it cannot act as a source for diffusion of Ge
into the heterostructure.  However, it is to our knowledge
not yet understood why the diffusion of Au underneath the
Ni grains at later stages of annealing (when a large amount
of Ge already diffused out of Ni) results in a strong
increase of the contact resistance.


\section{\label{sec:annmech}Summary of annealing mechanism}

In this Section we use the results from the previous two
Sections, together with established results from the
literature, for giving a qualitative description of the
formation of an ohmic contact to a 2DEG in a GaAs/${\rm
Al}_{x}{\rm Ga}_{1-x}{\rm As}$ heterostructure. It is
remarkably similar to the annealing mechanism as described
by Kuan \textit{et al.} \cite{Kuan1983} for contacts to
\mbox{n-GaAs}. In the initial stages of the annealing
process (already during the heating of the sample) Au and
Ge segregate, and most Ge forms a new phase with the Ni. At
the same time, these Ge-rich Ni grains move to the wafer
surface due to a wetting effect \cite{Relling1988}, which
results in the situation that the wafer surface is covered
with neighboring Au and ${\rm Ni_{x}Ge}$ \cite{NiGe}
grains. There is evidence that for thin metallization
layers ($\sim$100~nm) this process already occurs well
below the bulk melting temperature (363~$^{\circ}$C) of the
eutectic AuGe phase \cite{Relling1988}.

Next, at higher temperatures, both the Au-rich and Ni-rich
grains penetrate into the heterostructure by solid phase
inter diffusion, compensated by a back flow of As and Ga.
Our EDX results confirm that Ga mainly flows into Au, and
As mainly into Ni-rich grains. This concerns the formation
of new material phases. In several earlier studies
\cite{Kuan1983,Zwicknagl1986,Procop1987,Bruce1987,Relling1988,Weizer1988,Lumpkin1996}
these phases have been identified as AuGa alloys and phases
close to ${\rm Ni_{2}GeAs}$. These phases penetrate only
tens of nm below the original wafer surface for typical
annealing conditions \cite{Kuan1983,Bruce1987}.

At the same time, there is diffusion of \textit{atomic} Ge,
Ni and Au (at similar concentrations) into the
heterostructure, which penetrates deeper
\cite{Heiblum1982,Zwicknagl1986,Higman1986,Shappirio1987,Saravanan2008}.
In particular, Ge diffuses out of the Ni-rich grains into
the ${\rm Al}_{x}{\rm Ga}_{1-x}{\rm As}$ layers, and a good
ohmic contact is formed when the ${\rm Al}_{x}{\rm
Ga}_{1-x}{\rm As}$ layers are sufficiently doped with Ge
all the way up to the 2DEG. While this is progressing, the
Au-rich grains start to expand underneath the Ni-rich
grains \cite{Kuan1983,Lumpkin1996}, which have the lowest
contact resistance with the doped ${\rm Al}_{x}{\rm
Ga}_{1-x}{\rm As}$ layer since they were the dominant
supplier of Ge. The expansion of the AuGa grains is
possibly due to the relatively low activation energy for
out diffusion of Ga into Au \cite{Kulkarni1988b} (while the
\mbox{Al-As} binding energy is relatively high
\cite{Zwicknagl1986}). This latter process increases the
interface resistance between the metallization on the
surface and the doped ${\rm Al}_{x}{\rm Ga}_{1-x}{\rm As}$
layer. Thus, the formation of an optimal contact is a
competition between these two processes.

The in-diffusion of Ge lowers the contact resistance for
two reasons. 1) The full ${\rm Al}_{x}{\rm Ga}_{1-x}{\rm
As}$ region between the surface and the 2DEG becomes a
highly doped region with a reasonably low bulk resistivity.
2) The Ge doping in this region makes the Schottky barrier
between the doped semiconductor and the surface
metallization very thin (the barrier height is probably not
changing significantly \cite{Schottky}), up to the point
where its series contribution to the contact resistance is
small. The total contact resistance is then dominated by
doped ${\rm Al}_{x}{\rm Ga}_{1-x}{\rm As}$ region, giving
linear transport characteristics (a similar effect occurs
for contacts to \mbox{n-GaAs} due to spreading resistance
below the contact \cite{Braslau1981}).

As said, it is not yet well established which processes are
responsible for the resistance increase upon over
annealing. The fact that over annealing with 2DEG samples
and \mbox{n-GaAs} samples \cite{Kuan1983} occurs
qualitatively in a very similar manner (and also at similar
annealing times and temperatures) is a first indication
that it is due to a process near the interface with
metal-rich phases on the surface, rather than a process at
the depth of the 2DEG or the edge of a contact. Further,
our results now show that the resistance increase for 2DEG
samples is also correlated with the expanding AuGa grains
below the Ni-rich grains. Various authors have suggested
that the increasing contact resistance that is associated
with over annealing may be due to a large number of
vacancies just below the metal-rich phases near the surface
\cite{Zwicknagl1986,Bruce1987,Messica1995} (but others
suggested it was due to excessive in-diffusion of Ni
\cite{Ogawa1980,Bruce1987}). These mainly result from
out-diffusion of Ga into the Au-rich grains (which indeed
results in a very stable AuGa phase near the original wafer
surface \cite{Ogawa1980,Relling1988,Weizer1988}). These
vacancies occur in particular when there is no (longer) Ge
diffusion into these vacancies. One should note, however,
that with \mbox{n-GaAs} an increasing contact resistance
was also observed without an expansion of the AuGa grains
below the Ni-rich grains \cite{Bruce1987}, but this does
not rule out that an increasing number of vacancies is
responsible for over annealing.

Finally we remark that both the Ni-rich and Au-rich grains
are probably important for rapid annealing at relatively
low temperatures. The Ni-rich grains act as the supplier of
Ge. The presence of Au grains may be important since it
rapidly results in a large number of Ga vacancies. This
probably enhances the in-diffusion of Ge. It was for
example also observed that the creation of such vacancies
near the surface, enhances the diffusion of Si dopants from
the doping layer (much deeper into the material)  into
neighboring layers \cite{Shappirio1987}.


\section{\label{sec:model}Diffusion model}

\begin{figure}[h!]
\includegraphics[width=1.0\columnwidth]{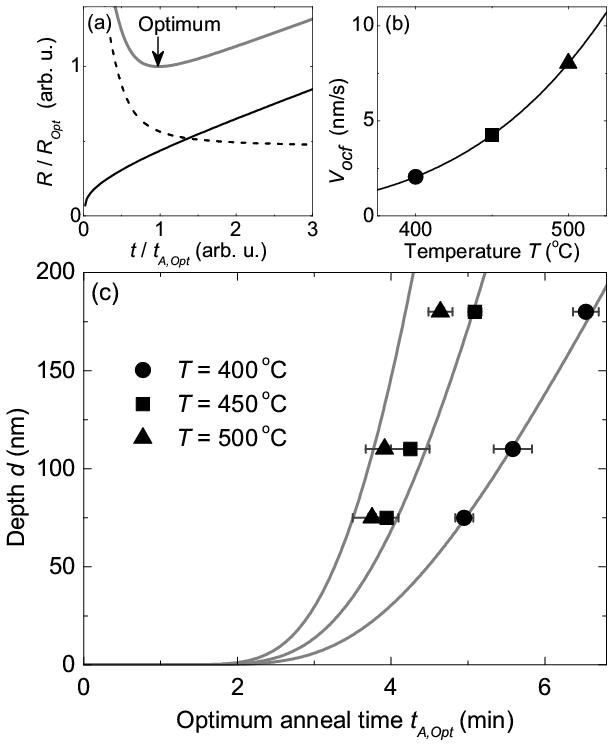}

\caption{(a) Model for the resistance of an ohmic contact
as a function of annealing time at constant temperature.
The resistance $R_{Ge}$ of the ${\rm Al}_{x}{\rm
Ga}_{1-x}{\rm As}$ layers (dashed line) decreases in time
due to increased Ge doping. The interface resistance
$R_{if}$ between the surface metallization and the Ge-doped
${\rm Al}_{x}{\rm Ga}_{1-x}{\rm As}$ layers (solid black
line) increases in time due to a decreasing Ni-grain--${\rm
Al}_{x}{\rm Ga}_{1-x}{\rm As}$ interface area. The time
where the sum of these two resistances (gray solid line)
shows a minimum defines the optimum annealing time
$t_{A,Opt}$. (b) Effective velocity of optimal contact
formation $v_{ocf}$ as a function of temperature
(Eq.~\ref{Eq:Vocf}), plotted for parameters that give the
best fit in (c). (c) Optimal annealing times as the 2DEG
depth and annealing temperature is varied (same
experimental data as in Fig.~\ref{Fig:TimeTempDepth}c). The
solid gray lines (left to right for 500, 450 and
400~$^{\circ}$C) represent fits using the model of
Eqs.~\ref{Eq:tAopt} and \ref{Eq:Vocf} (see text for
details).} \label{Fig:TimeDepthModel}
\end{figure}

We use the above description to construct a model that
predicts the optimal annealing time for a given annealing
temperature and 2DEG depth $d$. The contact resistance is
then the series resistance of the resistance of the
Ge-doped ${\rm Al}_{x}{\rm Ga}_{1-x}{\rm As}$ region
($R_{Ge}$) and the interface resistance between the surface
metallization and this Ge-doped ${\rm Al}_{x}{\rm
Ga}_{1-x}{\rm As}$ layer ($R_{if}$). For both we consider
the average over the full contact area. We will first
assume an anneal temperature $T$ that is constant in time.
We model the resistance of the Ge-doped ${\rm Al}_{x}{\rm
Ga}_{1-x}{\rm As}$ region using the result from work on
\mbox{n-GaAs} that the contact resistance is inversely
proportional to the doping concentration
\cite{Braslau1981}. Thus, we assume that
      \begin{equation} \label{Eq:Rge}
           R_{Ge} \propto \int \frac{1}{C(z)/C_0} dz ,
      \end{equation}
where $C(z)/C_0$ is the local Ge concentration at depth $z$
as in Eq.~\ref{Eq:FickDiffusion}, and where the integral
runs from the depth of the Au and Ni grains to the depth of
the 2DEG. The behavior of this equation is that $R_{Ge}$
first rapidly decreases, and then curves off to saturate at
a level that is proportional to $d$ (dashed curve in
Fig.~\ref{Fig:TimeDepthModel}a).


To model $R_{if}$, we assume that the increase in
resistance for over annealed contacts is related to the
decrease in Ni-grain--${\rm Al}_{x}{\rm Ga}_{1-x}{\rm As}$
interface area. Imagine, for simplicity, a single, square
shaped Ni-rich grain with area $A_{Ni} = L_{Ni}^2$. We
model the reduction of this area as a sideways diffusion
process of Au, again with a time-dependence as simple
diffusion analogues to Eq.~\ref{Eq:FickDiffusion}. The
length of a side is then reduced as $L_{Ni}(t) \approx L_0
- 2 \sqrt{4 D_{Au}t}$, where $L_0$ is the initial grain
size, and $D_{Au}$ the diffusion constant for this process,
such that
       \begin{equation} \label{Eq:Rif}
            R_{if} \propto \frac{1}{(L_0 - 2 \sqrt{4 D_{Au}t}\,)^{2}} .
       \end{equation}
For a very wide parameter range, this model gives that
$R_{if}$ increases more or less linearly in time (solid
black curve in Fig.~\ref{Fig:TimeDepthModel}a). A
resistance increase that is much stronger than linear only
sets in when the total interface area approaches zero, when
the contact is already strongly over annealed. The total
contact resistance is the sum of $R_{Ge}$ and $R_{if}$
(gray solid curve Fig.~\ref{Fig:TimeDepthModel}a), and the
optimal annealing time is then defined as the time where
this sum shows a minimum value.

We can reduce the number of fitting parameters for this
modeling to only two with the following approach. For
$R_{Ge}$ in Eq.~\ref{Eq:Rge}, we assume parameters where
$R_{Ge}$ saturates at a value below, but on the order of
the optimal contact resistance $R_{opt}$. We also assume
that this saturation occurs in a time scale on the order of
a few times the optimal annealing time. For $R_{if}$ in
Eq.~\ref{Eq:Rif}, we assume that it has a value below
$R_{opt}$ for $t=0$, and that it increases more or less in
a linear fashion to a value of order $R_{opt}$. This
increase should take place in a time scale on the order of
the optimal annealing time. Numerically investigating this
model then shows that it has for a very wide parameter
range the behavior that the increase of optimal annealing
time $t_{A,Opt}$ with increasing 2DEG depth $d$ is close to
linear. We can express this using an effective velocity for
optimal contact formation $v_{ocf}$,
       \begin{equation} \label{Eq:tAopt}
            t_{A,Opt} = d / v_{ocf}.
       \end{equation}
Furthermore, numerical investigation of the temperature
dependence shows that $v_{ocf}$ behaves according to
      \begin{equation} \label{Eq:Vocf}
           v_{ocf}(T) = v_0~{\rm exp}(-\frac{E_a}{k_B T})
      \end{equation}
when the diffusion processes that underlie Eq.~\ref{Eq:Rge}
and Eq.~\ref{Eq:Rif} are both thermally activated with a
similar activation energy $E_a$. We can now fit this model
to our experimental data only using Eq.~\ref{Eq:tAopt} and
Eq.~\ref{Eq:Vocf}, such that we only have $v_0$ and $E_a$
as fitting parameters. In doing so, we take again into
account that the temperature $T(t)$ is not constant during
annealing, and use again profiles as in
Fig~\ref{Fig:TimeTempDepth}a.

The results of this fitting are presented in
Fig.~\ref{Fig:TimeDepthModel}c, and $v_{ocf}$ as a function
of temperature for these fitting parameters ($E_a = 0.6 \;
{\rm eV}$ and $v_0 = 7.6 \cdot 10^{-5} \; {\rm m/s}$) is
plotted in Fig.~\ref{Fig:TimeDepthModel}b. While it is a
crude model, the fits are very reasonable, showing that the
model is useful for predicting optimal annealing times.
Furthermore, the value for $E_a$ is a realistic number
\cite{Sarma1984,Kulkarni1988a,Kulkarni1988b}. Our model
also predicts that the minimum value of the resistance that
can be achieved for optimally annealed contacts increases
with increasing 2DEG depth. We did not observe such a clear
trend, probably because the resistance of optimal contacts
is so low that one needs to include contributions from 2DEG
square resistance around and underneath the contact when
evaluating absolute values (further discussed below).


\section{\label{sec:shape}Contact-shape dependence}

Our model for the annealing mechanism implies that optimal
contacts have a rather uniform Ge concentration throughout
the ${\rm Al}_{x}{\rm Ga}_{1-x}{\rm As}$ layers, and that
this results in a value for $R_{Ge}$ of order 10~$\Omega$.
This implies that the bulk resistivity in the doped
Ge-doped ${\rm Al}_{x}{\rm Ga}_{1-x}{\rm As}$ layer is
around 4~$\Omega {\rm m}$. In turn, this implies that
in-plane electron transport under an optimal contact from
the metallization on the surface to 2DEG on the side of the
contact still mainly takes place in the original 2DEG
layer. If the square resistance $R_{\Box}$ for transport in
the original 2DEG layer below the contact does not strongly
increase during annealing, and if it is smaller than the
contact resistance, this also implies that the resistance
of optimal contacts should be inversely proportional to the
contact area. Thus, measuring whether the contact
resistance depends on contact area or on the circumference
of a contact can give further insight in the annealing
mechanism and contact properties.

We carried out such a study, by varying the shape of
contacts. All results that we discussed up to here were
obtained with square contacts with an area $A$ of
0.04~${\rm mm^2}$ and a circumference $C_{L}=4L$ of 0.8~mm
(on the side of a Hall bar). For the dependence on contact
shape, we measured various sets where we varied the
circumference $C_{L}$ while keeping the area constant at
0.04~${\rm mm^2}$, and various sets where we varied the
area while keeping the circumference constant at 0.8~mm. We
varied the shape from smooth circular shape to square
shapes with a zig-zag edge at the 50 micronscale, to avoid
getting too much resistance contribution from square
resistance of 2DEG right next to a contact (for these
devices we used electron-beam lithography). The study only
used wafer~A. All contacts were fabricated and annealed in
one single batch to ensure that it is meaningful to compare
the values of contact resistance.

For this study, we inject again current into the contact
that is measured, and extract the current using another
contact. However, the dependence on contact shape can only
give an unambiguous result if the resistance from each side
of the studied contact to the place in the 2DEG where the
current is extracted is sufficiently similar. This can be
achieved by making the distance between the contacts larger
than the size of the contacts. Thus, we now fabricated
contacts in the middle of 2~mm by 3~mm cleaved wafer pieces
(two rows of four contacts, with center-to-center distance
between rows 1~mm, and center-to-center distance between
contacts within a row 0.6~mm). Using four different
contacts for a 4-terminal measurement on the 2DEG (with the
current biased from one row to the other) gives on these
samples indeed low values around 8~$\Omega$, in reasonable
agreement with the value of the 2DEG square resistance
$R_{\Box}$ of about 20~$\Omega$. Contact resistance values
were again determined in a current-biased 4-terminal
configuration, with two terminals connected to the bond
wire on the contact, the second current terminal on the
opposite contact in the other row, and the second voltage
terminal on a neighboring contact in the same row.

\begin{figure}[h!]
\includegraphics[width=1.0\columnwidth]{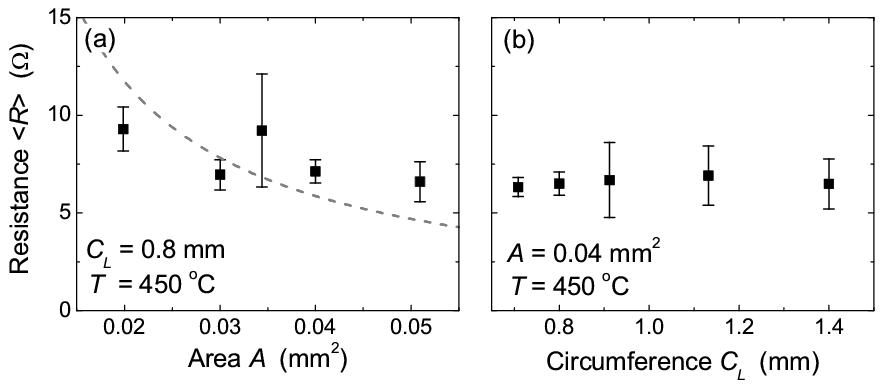}
\caption{Contact resistance $\langle R \rangle$ as a
function of (a) contact area $A$ for constant circumference
$4L$ and (b) contact circumference $C$ for constant area
$A$. The error bars here represent the standard deviation
from measuring $R$ on 8 identical contacts. The dashed line
in (a) is a fit using $\langle R \rangle \propto 1/A$.}
\label{Fig:AreaCircumference}
\end{figure}

On contacts that are not annealed, we can observe a tunnel
current, as expected for Schottky barriers. Here, the
effective resistance is inversely proportional to area. For
optimally annealed contacts, we found that the contact
resistance was independent on circumference, while only
showing a weak dependence on area (weaker than inversely
proportional to area), see
Fig.~\ref{Fig:AreaCircumference}. The fact that the
dependence on shape does here not show a clear dependence
as $\langle R \rangle \propto 1/A$ agrees with the fact
that the $\langle R \rangle$ values are comparable to the
square resistance of the 2DEG, such that the latter gives a
significant contribution to the total contact resistance.
Fully understanding the contact resistance then requires
incorporating all square resistance contributions from
underneath and around the 2DEG. Since we found it
impossible to estimate these effects with a small error
bar, we tried to demonstrate a clear dependence on area by
measuring slightly under annealed contacts instead.


On two sets of under annealed contacts on wafer~A, where we
used shorter anneal times than $t_{A,Opt}$ (average contact
resistance of 30~$\Omega$ and 500~$\Omega$), we found
(within error bar) no dependence on area or circumference.
We can only explain this result if we assume that the 2DEG
square resistance underneath the contact is significantly
increased (to values comparable to the total observed
contact resistance) for under annealed contacts. This
probably results from the in-diffusing Ge (and atomic Au
and Ni \cite{Shappirio1987,Saravanan2008}), which already
introduces strain and scatter centers in the 2DEG layer
before optimal contact conditions are reached. For optimal
annealed contacts (here with total resistance of typically
7~$\Omega$, independent of circumference), the square
resistance underneath a contact must have returned to a low
value of order 10~$\Omega$. Apparently, the resistance
increase due to strain and scatter centers is compensated
by increased Ge doping near the 2DEG layer.

The summary of this study is that the resistance of
annealed contacts never shows in a clear dependence on
circumference, and only a weak dependence on area for
optimal contacts. We can, nevertheless, draw the following
conclusions. For an optimal ohmic contact, it is
\textit{not} the case that electron transport between the
surface metallization and the surrounding 2DEG mainly
occurs at the edge of a contact. Instead, the full contact
area plays a role, and in-plane electron transport under an
optimal contact mainly takes place in the original plane of
the 2DEG. In addition, we find it impossible to evaluate
the absolute values of the contact resistance of our
devices with an accuracy within a factor 2, since the
resistance of an optimal contact has a contribution from
the square resistance underneath the contact, and its value
is influenced by the annealing process. Further, future
studies in this direction should consider that pressing the
bond wire (with a footprint of about 100$\times$100~${\rm
\mu m}^2$) onto the surface metallization may locally
disturb the contact properties, which can disturb a clear
dependence on contact shape.

We could therefore not study the property of our model that
the optimal contact resistance value should be proportional
to $d$. Instead, we should evaluate whether the enhanced
square resistance underneath a contact needs to be
incorporated in our model. We find that this is not needed
for the following reasons: for over annealing it does not
play a role, since we observe the same over-annealing
mechanism as observed on bulk \mbox{n-GaAs}. Optimally
annealed contacts occur when the square resistance
underneath the contacts has again low values of order
10~$\Omega$. Here we observe a weak area dependence and no
dependence on circumference, such that we can rule out that
the effect dominates the contact resistance. Thus, the only
effect is that it temporarily enhances the total contact
resistance by about a factor 2 while the annealing
progresses towards optimal contact conditions. Note that
this does not change the fact that lowering the contact
resistance in this phase still fully depends on further Ge
diffusion towards the 2DEG layer. Therefore, it only
slightly modifies how $R_{Ge}$ in Eq.~\ref{Eq:Rge}
decreases towards low values.


\section{\label{sec:conclu}Conclusions}

Summarizing, we have measured the zero-bias resistance of
annealed AuGe/Ni/Au ohmic contacts to a 2DEG as a function
of annealing time and temperature. We have thus obtained
optimal annealing parameters for three different
heterostructures where the 2DEG lies at a different depth
below the wafer surface. TEM images of several annealed
contacts provided further insight into the annealing
mechanism and the formation of a good ohmic contact.

Combining this information we have developed a model that
can predict the optimal annealing parameters for contacting
a 2DEG at a certain depth in a GaAs/${\rm Al}_{x}{\rm
Ga}_{1-x}{\rm As}$ heterostructure. The model assumes two
competing processes: 1) Diffusion of Ge into the
heterostructure lowers the contact resistance, and results
in linear transport characteristics. 2) At longer annealing
times, Au-rich phases diffuse in between the
heterostructure and Ni-rich phases at the wafer surface.
The associated increase in contact resistance is probably
due to subsequent diffusion of Ga into this Au-rich phase,
since this increases the number of Ga vacancies in the
heterostructure near the metallization on the surface. The
competition between these two processes results in a
mechanism where the optimal annealing time (for a process
at constant annealing temperature) is proportional to the
depth of the 2DEG below the surface, and the speed of this
process has thermally activated behavior. This model should
have predictive power for many heterostructures, as long as
the temperature of the samples as a function of time during
the annealing process is known. Our study of how the
contact resistance depends on the shape of the contact
confirmed that the full contact area plays a role in
electron transport between the metallization on the surface
and the 2DEG.

Our model may become invalid for systems with a very deep
2DEG, since $R_{if}$ (Eq.~\ref{Eq:Rif}) is expected to
increase more rapidly at long annealing times, possibly
resulting in non-ohmic behavior. Our results suggests that
for solving this problem the focus should be at maintaining
sufficient contact area between Ge-rich Ni grains and the
Ge-doped ${\rm Al}_{x}{\rm Ga}_{1-x}{\rm As}$ layer at long
annealing times. This can possibly be engineered by
changing the layer thickness, order and composition of the
initial AuGe/Ni/Au metallization
\cite{Zwicknagl1986,Procop1987,Bruce1987,Shappirio1987,Relling1988,Rai1988,Raiser2005},
or by including a Pt, Nb or Ag layer below the top Au layer
that suppresses the intermixing of this Au with layers at
the wafer
surface\cite{Lee1981,Zwicknagl1986,Shappirio1987,Jin1991,Messica1995}
(uniform Ni/Ge/As layers been reported \cite{Rai1988}).
Alternatively, one can reduce the depth of the 2DEG by
etching before deposition of AuGe/Ni/Au (up to the point
where this results in depletion of the 2DEG).

\section*{ACKNOWLEDGMENTS}
We thank B.~H.~J.~Wolfs, M.~Sladkov and S.~J.~van der Molen
for help and valuable discussions. Further, we acknowledge
the Dutch Foundation for Fundamental Research on Matter
(FOM), the Netherlands Organization for Scientific Research
(NWO), and the German programs DFG-SFB 491 and
BMBF-nanoQUIT for funding.











\end{document}